\documentclass[pra,twocolumn,showpacs,floatfix]{revtex4-1}
\usepackage[utf8]{inputenc}
\usepackage[T1]{fontenc}
\usepackage{ae}
\usepackage{amsmath,amssymb,amsfonts}
\usepackage{bm,bbm}
\usepackage{bbm}
\usepackage{units}
\usepackage{graphicx}
\usepackage{dcolumn}
\usepackage{float}
\usepackage{mathrsfs}

\newcommand{\abs}[1]{\left| #1 \right|} 					
\newcommand{\ii}{\mathrm{i}}							
\newcommand{\ee}[1]{\mathrm{e}^{#1}}						
\newcommand{\dd}{\mathrm{d}}							
\renewcommand{\vec}[1]{\bm{#1}}							
\newcommand{\vectornorm}[1]{\left|\left| #1 \right|\right|}			

\newcommand{\braket}[2]{\left< #1 \vphantom{#2} \right|
 \left. #2 \vphantom{#1} \right>} 						
\newcommand{\matrixel}[3]{\left< #1 \vphantom{#2#3} \right|
 #2 \left| #3 \vphantom{#1#2} \right>} 						
\newcommand{\ket}[1]{\left| #1 \right>}						
\newcommand{\bra}[1]{\left< #1 \right|}						
\newcommand{\del}[2]{\frac{\partial #1}{\partial #2}}				
\DeclareMathOperator{\Real}{Re}							
\DeclareMathOperator{\Imag}{Im}							
\DeclareMathOperator{\Tr}{Tr}							

\DeclareMathOperator{\erfc}{erfc}

\begin{document}

\title{Collisions of anisotropic two-dimensional bright solitons in
  dipolar Bose-Einstein condensates}

\author{R\"udiger Eichler}
\email{ruediger.eichler@itp1.uni-stuttgart.de}
\author{Damir Zajec}
\email{zajec@itp1.uni-stuttgart.de\\(Both authors contributed equally to this work)}
\author{Patrick K\"oberle}
\author{J\"org Main}
\author{G\"unter Wunner} 
\affiliation{Institut f\"ur Theoretische Physik 1, Universit\"at Stuttgart,
70550 Stuttgart, Germany}

\date{\today}

\begin{abstract}
  We investigate the coherent collision of anisotropic
  quasi-two-dimensional bright solitons in dipolar Bose-Einstein
  condensates. Our analysis is based on the extended Gross-Pitaevskii
  equation, and we use the split-operator method for the grid
  calculations and the time-dependent variational principle with an
  ansatz of coupled Gaussian functions to calculate the time evolution
  of the ground state. We compare the results of both approaches for
  collisions where initially the solitons are in the repelling
  side-by-side configuration and move towards each other with a
  specific momentum. We change the relative phases of the condensates,
  and introduce a total angular momentum by shifting the solitons in
  opposite direction along the polarization axis. Our calculations
  show that collisions result in breathing-mode-like excitations of
  the solitons.
\end{abstract}
\maketitle

\section{Introduction}
\label{sec:introduction}

Bose-Einstein condensates (BECs) of magnetic atoms have attracted much
attention since their experimental realization with $^{52}$Cr atoms
\cite{Griesmaier2005,*Beaufils2008}. Recently, the creation of
condensates of $^{164}$Dy \cite{Lu2010,*Lu2011_2} and
${}^{168}\mathrm{Er}$ \cite{Aikawa_2012} atoms with much larger
magnetic moments than $^{52}$Cr have also been reported. Furthermore,
there has been fast progress towards the condensation of polar
molecules with electric dipole moments \cite{Ni_2008}, where the
dipole-dipole interaction (DDI) is even more dominant. A review of the
physics of dipolar bosonic quantum gases has recently been given by
Lahaye \textit{et al.} \cite{Lahaye2009}. The features of the DDI
being a non-local long-ranged and anisotropic interaction give rise to
a variety of new effects. One example is the creation of solitary
waves, where in analogy to nonlinear optics the effects of dispersion
and nonlinearity may cancel each other. This leads to a condensate
with a shape constant in time. The experimental realization of
one-dimensional solitons in self-attractive BECs of $^{7}$Li atoms has
been reported \cite{Khaykovich2002,*Strecker2002}. Tikhonenkov
\textit{et al.}\ have theoretically predicted 2D solitons
\cite{Tikhonenkov2008_2} and K\"oberle \textit{et al.}\ have proposed
a realistic experimental setup for the creation of a 2D soliton
\cite{Koeberle2012}. An exciting aspect of multidimensional solitons
is their anisotropic nature, based on the in-plane polarization of the
dipoles of such solitons. 2D solitons have already been studied using
a variational ansatz with a single Gaussian and
with coupled Gaussian functions \cite{Eichler2011}. Adhikari
\textit{et al.}\ have recently investigated axially symmetric and
vortex solitons on a one-dimensional optical lattice
\cite{Adhikari_2012}. Note that in contrast to systems with harmonic
traps, where the density distribution in the trap direction is an
approximate Gaussian, systems in an optical lattice will have an
exponential density distribution.

The collision of axially symmetric bright 2D solitons has been studied
by Pedri \textit{et al.} \cite{Pedri2005} and Adhikari \textit{et al.}
\cite{Adhikari_2012}. Pedri \textit{et al.} investigated a system with
dipoles aligned parallel to the harmonic trap, while Adhikari
\textit{et al.} used an optical lattice instead. In both cases,
the sign of the DDI has to be inverted by fast rotation of the
orientation of the dipoles \cite{Giovanazzi2002}. The resulting
interaction energy becomes
$U_{\rm{d}}(\bm{R})=-\alpha(3\cos^2{\vartheta}-1)/\bm{R}^3$, where
$\vartheta$ is the angle between the polarization axis and
$\bm{R}=\bm{r}-\bm{r'}$. The factor $\alpha$ can continuously be
changed from $-1/2$ to $1$. This provides the possibility to change
the dipolar interaction from attractive to repulsive.

In addition to the analysis of the collision of 2D solitons, Young
\textit{et al.}\ \cite{Young2011} have investigated the collision of
one-dimensional bright and vortex solitons. The investigations in
\cite{Tikhonenkov2008_2,Koeberle2012,Eichler2011} concentrated on the
creation and the stability of 2D solitons with respect to small
perturbations. However, one important property of solitons is that
their shape is constant in time even when they are moving. Therefore,
the collision of two solitons is an adequate scenario for the
investigation of soliton dynamics far beyond small excitations. The
influence of the nonlinear contact interaction and the DDI are of
particular interest in such calculations.

As mentioned above, the creation of a BEC of magnetic atoms has been
realized with a variety of species. Our results are valid for all
dipolar systems, but we will add the corresponding values for a system
with $20\,000$ $^{52}$Cr-atoms per soliton in parentheses.

At sufficiently low temperatures, the dynamics of a Bose-Einstein
condensate can be described by the extended Gross-Pitaevskii equation
(GPE) which in atomic units and with particle-number scaling
\cite{Koeberle09a} reads

\begin{align}
  H(t) \Psi(\vec{r},t) =& \left( -\Delta + V_{\mathrm{har}} +
      V_{\mathrm{sc}} + V_{\mathrm{d}} \right) \Psi (\bm{r},t)\nonumber\\
    =& \mathrm{i} \partial _{t}\Psi (\bm{r},t)\,,\\
    \text{with}\quad 
    V_{\mathrm{har}} =& \gamma_y^2 y^2 \,,\quad
    V_{\mathrm{sc}} = 8\pi a \left| \Psi(\vec{r},t) \right|^2 \,,\nonumber\\
    V_{\mathrm{d}} =& \int \mathrm{d}^{3}r^{\prime }\,\frac{%
      1-3\cos ^{2}\vartheta}{\left\vert
        \bm{r}-\bm{r}^{\prime }\right\vert ^{3}}\left\vert \Psi
      (\bm{r}^{\prime
      },t)\right\vert ^{2} \,.\nonumber
\end{align}
Here $a$ is the scattering length and $\Psi$ designates the mean-field
wave function. The dipoles are polarized along the $z$-axis, so that
$\vartheta$ is the angle between the $z$-axis and the vector
$\bm{r}-\bm{r}'$. We choose the $y$-direction as the axis of
confinement perpendicular to the polarization axis where
$\gamma_y=20\,000$ ($420$ Hz), while the condensate is free in
$x$- and $z$-direction. All simulations deal with condensates of low
densities, and only a small period of time in which the two condensates
merge to one transient condensate with higher density. This means that
we do not need to take a three-body-loss term \cite{Koeberle2012} into
account, as the resulting absorption images ($|\psi|^2$ integrated
along the $y$-axis) would only be slightly affected. We checked this
assumption for the calculation of the collision without difference in
phase and without angular momentum which up to the time of $t=0.06$
($t=0.001$ corresponds to $15$ ms) resulted only in a loss of about
$5.5\%$ of the particles.

As has been shown in
\cite{Eichler2011}, solitons only exist in a certain range of values
of the scattering length, which can be tuned by the use of Feshbach
resonances \cite{Chin2010}. For too large values, the condensate will
disperse, while too small values lead to the collapse of the condensate. 
In the following the scattering length is chosen to be $0.14$ ($12.7 a_\mathrm{B}$, where 
$a_\mathrm{B}$ is the Bohr radius) .

\section{Numerical Approach}
\label{sec:numerical approach}

The main theoretical task for the grid calculations is how to apply
the time evolution operator $U=e^{-iHt}$ on a state $\ket{\psi}$. For
this, one splits $U$ symmetrically by using the
Baker-Campell-Hausdorff formula \cite{Feit_1982}

\begin{align}
 U(\Delta t) & = e^{-iH\Delta t} = e^{-i(T+V)\Delta t} \nonumber \\
  & \approx e^{-i\frac{1}{2}T\Delta t}e^{-iV\Delta t}e^{-i\frac{1}{2}T\Delta t},
\label{eq:BCH} 
\end{align}
where $V = V_{\rm{har}} + V_{\rm{sc}} + V_{\rm{d}}$.
One projects the action of the approximated time evolution operator
on the basis of the position operator and makes use of the possibility
to insert $\int \text{d} \nu \ket{\nu}\bra{\nu} = 1$:

\begin{align}
 \psi(\vec{r}, t + \Delta t) = & \bra{\vec{r}} U(\Delta t) \ket{\psi} \nonumber \\
 \nonumber = & \int \text{d}^{3}p^{\prime} \text{d}^{3}r^{\prime} \text{d}^{3}p \bra{\vec{r}} 
 e^{-i \frac{p^{2}}{2} \Delta t} \ket{\vec{p}^{\prime}} \\
 \nonumber & \bra{\vec{p}^{\prime}} e^{-i V(\vec{r})\Delta t} \ket{\vec{r}^{\prime}}
  \bra{\vec{r}^{\prime}} e^{-i \frac{p^{2}}{2}\Delta t} 
  \ket{\vec{p}} \braket{\vec{p}}{\psi} \\
 \nonumber = & \frac{1}{\sqrt{2\pi}^9} 
 \int \text{d}^3p' \text{d}^3r' \text{d}^3p e^{i\vec{r}\vec{p}'} e^{-i \frac{p'^2}{2}\Delta t} \\
 & e^{-i\vec{p}'\vec{r}'} e^{-i V(\vec{r}')\Delta t} e^{i \vec{r}' 
 \vec{p}} e^{-i \frac{p^2}{2}\Delta t} \tilde{\psi}(\vec{p}).
 \label{eq:time_evolution}
\end{align}
The structure of \eqref{eq:time_evolution} suggests the following
algorithmic procedure:
\begin{itemize}
 \item Fourier transform of $\psi(\vec{r})$ in order to obtain $\tilde{\psi}(\vec{p})$
 \item Multiply by $e^{-i \frac{p^2}{2}\Delta t}$
 \item Inverse Fourier transform to real space
 \item Multiply by $e^{-i V(\vec{r})\Delta t}$
 \item Fourier transform to momentum space
 \item Multiply by $e^{-i \frac{p^2}{2}\Delta t}$
 \item Inverse Fourier transform to real space
\end{itemize}
The potential $V$ consists of the harmonic potential, the scattering
potential and the DDI potential. The scattering potential and the DDI
potential have to be calculated at each time step. The latter can be
evaluated by means of the convolution theorem, which results
in two more Fourier transforms:

\begin{equation}
 \Phi_{\mathrm{dd}}(\vec{r}) = \frac{4\pi}{3}\mathscr{F}^{-1}\left\{\left(\frac{3k_z^2}{k^2}-1\right)
 \mathscr{F}\{|\psi(\vec{r})|^2\}\right\}.
\label{eq:interaction_potential}
\end{equation}
Here $k$ and $k_z$ denote the momentum and the momentum in
$z$-direction, respectively. Altogether, we have to perform six
Fourier transforms for each time step. Note that the first and last
Fourier transforms described in the algorithmic procedure of the time
evolution are only necessary if one is interested in physical
quantities whose evaluation requires the wave function in real space.

For the simulations, the spatial domain was 
discretized with up to $512 \times 128 \times 512$
grid points. Since this scheme is numerically very demanding, it has 
been implemented for graphics processing units (GPUs) using CUDA, 
enabling a very high degree of parallelization.
Using the Tesla C2070 improves the performance of our algorithm by a 
factor of about 80 for double precision in comparison to 
the corresponding C algorithm using the well known
FFTW library for computing the discrete Fourier transform on a 
IBM System x3400 with a Quad-Core Intel Xeon Processor 
E5430 (2.66GHz 12MB L2 1333MHz 80w) and 4 x 4GB PC2-5300 CL5 ECC DDR2 
Chipkill Low Power FBDIMM 667MHz.

To investigate the coherent collision of solitons we have applied the
following procedure. The first step is the computation of the ground
state of one condensate using the split-operator method with imaginary
time evolution ($t = -i \tau)$. Afterwards we double the size of the
grid in the $x$-direction and place two solitons in the repelling
side-by-side configuration. The distance between the condensates is
chosen such that they do not feel the mutual dipole-dipole
interaction. To introduce momentum in the system, we multiply the left
hand-side of the wave function by a plane wave $e^{ikx}$ (for the
soliton moving to the right) and the right hand-side by $e^{-ikx}$
(for the soliton moving to the left), respectively.

\section{Time-dependent variational ansatz}
\label{sec:time-depend-vari}

Variational calculations using coupled Gaussian wave packets (GWPs)
have shown to be a full-fledged alternative to numerical grid
calculations for the calculation of ground states of dipolar BECs
\cite{Rau10,Eichler2011}. The applicability of such ansatzes to
dynamical simulations is a challenging task. The decisive extension of
the previous work \cite{Rau10,Eichler2011} is that additional
translational and rotational degrees of freedom are included in the
ansatz with coupled GWPs to describe the dynamics of the condensate
wave function. For the convenience of the reader we shortly review the
time-dependent variational principle (TDVP) in this section, and
subsequently apply it to the ansatz of coupled GWPs. We make use of
the TDVP in the formulation of McLachlan \cite{McLachlan1964a} where
$\phi$ is varied such that
\begin{align}
  I = \vectornorm{\mathrm{i} \phi - H \Psi(t)}^{2}\stackrel{!}{=}\min\,,		\label{eq:McLachlan_I}
\end{align}
and set $\phi \equiv \dot \Psi$ afterwards. The
wave function $\Psi$ is considered to be parametrized by the
variational parameters $\Psi = \Psi(\vec z(t))$. The minimization of
the quantity $I$ in Eq.~(\ref{eq:McLachlan_I}) leads to 
\begin{align}
  \braket{\del{\Psi}{\vec{z}}}{\ii\dot{\Psi}- H \Psi } &=
  0 \,,\label{eq:McLachlan_matrix_bracket}
\end{align}
which can be written in the short form
\begin{align}
K \dot{\vec{z}} &= -\ii
\vec{h}\,,\label{eq:McLachlan_2.32_Kdotz=-ih}
\end{align}
with the positive definite Hermitian matrix $K$. We use a linear
superposition of $N$ Gaussian wave packets (GWPs)
\begin{align}
  \label{eq:ansatz_tdvp_eom}
    \Psi &= \sum\limits_{k=1}^{N} \ee{-
      \left(
        \left(
          \vec x^T - \vec q^k
        \right)^T A^k
        \left(
          \vec x - \vec q^k
        \right) - \ii\left(\vec p^k\right)^T
        \left(
          \vec x - \vec q^k
        \right) +  \gamma^k
      \right)}\nonumber\\
    &\equiv 
    \sum\limits_{k=1}^{N} g^k
    \,,
\end{align}
as an ansatz for the wave function in Eq.~(\ref{eq:McLachlan_I}). In
general, $A^k$ are $3\times 3$ complex matrices (determining the width
and the orientation of the GWP), $\vec p^k$ and $\vec q^k$ are
three-dimensional real vectors (representing momentum and center of
the GWP) and $\gamma^k$ are complex numbers (where the real part
stands for the amplitude and the imaginary part for the phase of the
GWP, respectively). In this work we will make use of the strong
confinement in one direction perpendicular to the dipole axis and
omit the translational and rotational degrees of freedom in $y$-direction
\begin{align}
  \label{eq:variational_reduction}
  A^k_{y \sigma} = A^k_{\sigma y} = 0\,,\quad
  p^k_y = 0\,,\quad q^k_y = 0\,,
\end{align}
with $\sigma= x,z\,.$ Inserting the ansatz
Eq.~(\ref{eq:ansatz_tdvp_eom}) in
Eq.~(\ref{eq:McLachlan_matrix_bracket}), sorting the result by powers
of $\vec x$ and identifying these terms with the coefficients of
a time-dependent effective harmonic potential
\begin{align}
  \label{eq:variational_V_eff}
  V_\mathrm{eff}^k = v_0^k + \vec v_1^k \vec x + \vec x V_2^k \vec x\,,
\end{align}
yields the equations of motion (EOM) for the variational parameters
\begin{subequations}
\begin{align}
  \dot A^k &= -4 \ii \left( A^k
  \right)^2 + \ii V^k_2  \,,\\
  \dot{\vec p}^k &= -\Real \vec v_1^k - 2 \Imag A^k \left( \dot{\vec
      q}^k - 2\vec p^k
  \right) - 2 \Real V_2^k \vec q^k  \,,\\
  \dot{\vec q}^k &= 2\vec p^k + \frac{1}{2} \left( \Real A^k
  \right)^{-1} \left( \Imag \vec v_1^k + 2 \Imag V_2^k \vec q^k
  \right)\,, \\
  \dot \gamma^k 
  &=
  2 \ii \Tr A^k - \ii \vec q^k V_2^k \vec q^k + 4\vec
  p^k A^k \vec q^k + \ii \left( \vec p^k
  \right)^2 \nonumber\\
  &\quad - \ii \vec q^k \dot{\vec p}^k - \ii \vec p^k \dot{\vec q}^k -
  2\vec q^k A^k \dot{\vec q}^k + \ii v_0^k    \,.
\end{align} \label{eq:var_EOM}
\end{subequations}
If we write  Eq.~(\ref{eq:McLachlan_2.32_Kdotz=-ih}) explicitly for
GWPs, the set of linear equations for $\dot{\vec z}$ can be rewritten
to one for the vector $\vec v$ containing the coefficients of $V^k_{\mathrm{eff}}$
\begin{align}
K \vec{v} &= \vec{r}\,,\label{eq:McLachlan_Kv=r}
\end{align}
for details, see \cite{Rau10a,Rau10b}. With the transformation given
in Appendix~\ref{sec:transformation-c-b} the EOM can now be integrated
with a standard algorithm such as Runge-Kutta, where
Eq.~(\ref{eq:McLachlan_Kv=r}) has to be solved at every time step.
The right-hand side vector $\vec r$ with the components
\begin{align}
  \label{eq:var_rhs_vec_r}
  \vec r^l &= \sum\limits_{k=1}^N\matrixel{g^l}{x_\alpha^m x_\beta^n
    V(\vec x)}{g^k}\,,
\end{align}
where $l=1,\dots,N;\; \alpha,\beta = 1,\dots,3; \; 0 \leq n+m \leq 2$,
contains integrals of the potentials in the GPE. It is one of the most
important advantages of the method that nearly all of these integrals
can be calculated analytically. However, the dipolar integral
$\matrixel{\Psi}{V_{\mathrm{d}}}{\Psi}$ can only be calculated
analytically for GWPs centered in the origin without the additional
translational degrees of freedom introduced in the ansatz
(\ref{eq:ansatz_tdvp_eom}). The analytical and numerical treatment of
the dipolar integral is presented in
Appendix~\ref{sec:vari-solut-dipol}.

The procedure for the calculations is as follows: At first the
equations of motion (\ref{eq:var_EOM}) for one soliton are integrated
in imaginary time, with the wave function being normalized after every
time step. Afterwards every GWP of the wave function is copied and the
resulting two solitons are positioned in the same way as given in
Sec.~\ref{sec:numerical approach}. Then for each GWP a corresponding
momentum $\vec p^k = \pm p^k_x \vec e_x$ is added. For this starting
configuration the EOM are finally integrated in real time.

\section{Results}
\label{sec:results}

\begin{figure}[tb]
 \includegraphics[width=1.0\columnwidth]{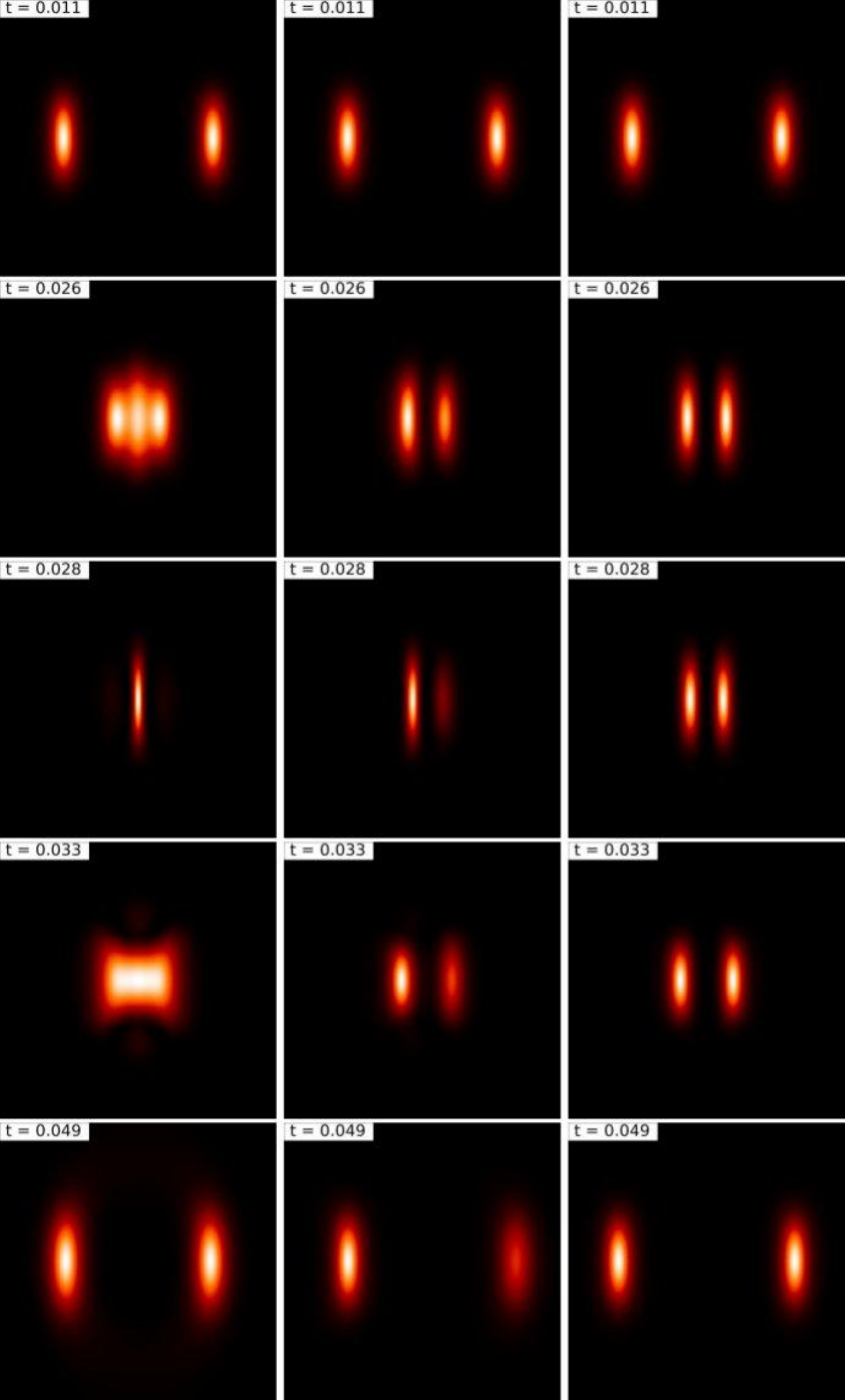}
 \caption{(Color online) Absorption images ($|\psi|^2$ integrated
   along the $y$-axis) of grid calculations for the collision of
   solitons. The value of the momentum for each soliton is $k=10$
   (velocity $v=127 \mu \rm{m} / s $) and the field of view is $1.4
   \times 1.4$ ($135 \mu \rm{m} \times 135 \mu \rm{m}$). All
   absorption images have been normalized to the maximum value. Left
   column: Absorption images for a collision without difference in
   phase. Middle column: Absorption images for a collision with a
   difference of $\phi=\pi/2$ in phase. Right column: Absorption
   images for a collision with a difference of $\phi=\pi$ in phase. }
\label{fig:results_10}
\end{figure}

In Fig.~\ref{fig:results_10} three grid calculations of colliding
solitons without angular momenta prepared in the way given above are
shown. For no phase difference constructive interference occurs and
the condensates merge and split up in two solitons again. Note that
the condensates after the split-up ($t = 0.049$, $t=0.001$ corresponds
to $15$ ms) have a larger spatial distribution than before ($t =
0.011$). 
This indicates that the transfer of kinetic energy to
internal energy has excited the solitons. This might either induce the
dispersal of the solitons or lead to breathing-mode-like oscillations.
The column in the middle shows a simulation with a difference of
$\phi=\pi/2$ in phase, resulting in a collision where the soliton on
the right eventually has a lower amplitude than the one on the left,
so that we do not have symmetric behavior anymore. The transfer of
kinetic energy is not as large as for $\phi=0$, resulting in only
slightly larger condensates at $t = 0.049$. In the case of a collision
with a difference of $\phi=\pi$ in phase we can see destructive
interference (column on the right), the solitons effectively repel
each other. The transfer of kinetic energy into internal energy is
once again smaller, corresponding to a just slightly larger condensate
at $t = 0.049$. The occurrence of the broken symmetry in $x$-direction
can be understood if one considers that a difference in phase of
$\phi=0$ and $\phi=\pi$ yields a wave function which is an
eigenfunction of the parity operator, in the sense of
$\Pi_{\pm}\Psi(\bm{r},t) = \pm\Psi(-\bm{r},t)$. A difference of
$\phi=\pi /2$ on the other hand does not result in an eigenfunction of
the parity operator, thus yielding an asymmetric dynamic of the
condensates.

\begin{figure*}[tb]
  \includegraphics[width=1.0\textwidth]{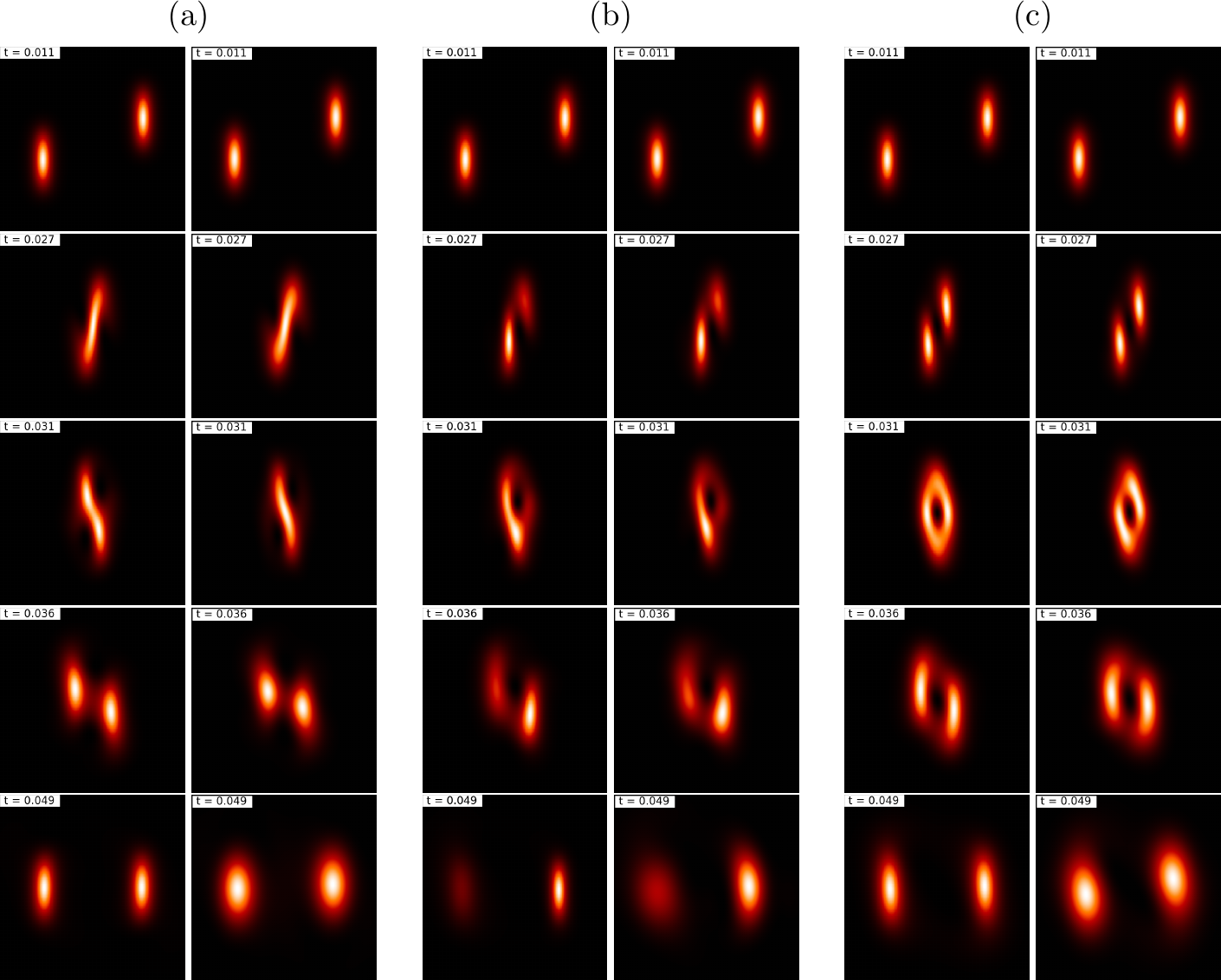}
  \caption{(Color online) Absorption images for grid calculations and
    the variational ansatz of the simulation of two colliding solitons
    with angular momentum. All absorption images have been normalized
    to the maximum value. The parameters are the same as given in
    Fig.\ \ref{fig:results_10}. The columns (a), (b) and (c) show
    calculations for a difference of $\phi=0$, $\phi=\pi /2$ and
    $\phi=\pi$ in phase, where the left column is the result of the
    grid calculations and the column on the right hand-side presents
    the results of the variational ansatz. For all three calculations
    six GWPs (three for each soliton) were used. The variational
    calculation is able to reproduce the transient ring-like structure
    during the collision for a difference of $\phi=\pi$ in phase and
    yields the correct results for the configuration at the end of all
    three calculations.}
\label{fig:collision_angMom_all}
\end{figure*}

In Fig.~\ref{fig:collision_angMom_all} we compare the results for grid
calculations and the variational ansatz for simulations, where we
shifted both condensates in opposite directions along the polarization
axis in order to introduce angular momentum. Both approaches are in
very good agreement with only slight differences, in particular for
times where both condensates merge, and when comparing the extensions
of the solitons at $t=0.049$. It is remarkable that a total number of
only six GWPs is sufficient to reproduce the structures of the grid
calculations and give the correct result for the configuration at the
end of all three simulations. The first case without a difference in
phase (Fig.~\ref{fig:collision_angMom_all}a) once again leads the
solitons to merge and split up afterwards, while a transient eddy-like
structure appears in the course of the collision. As in the case with
no angular momentum, the solitons either seem to disperse, or a
breathing-mode-like oscillation has been excited. The amount of
kinetic energy which has been transferred is lower than in the former
case, leading to condensates with smaller extension at $t = 0.049$
than their corresponding condensates in the simulation presented
above. A difference of $\phi=\pi/2$ in phase
(Fig.~\ref{fig:collision_angMom_all}b) shows a
similar behavior as in the case without angular momentum, resulting
once again in an asymmetric situation where after the collision the
condensates do not have the same amplitudes anymore. But for finite
angular momentum one may actually speak of a merged condensate at $t =
0.031$. Finally the collision with a difference of $\phi=\pi$
(Fig.~\ref{fig:collision_angMom_all}c) shows the
condensates effectively repelling each other, but in this case
introducing angular momentum leads to a transient ring-like structure.
The extension of the condensates after the collision is much larger
compared to the case with no difference in phase, which means that the
amount of transferred kinetic energy in internal energy is larger than
in the former case.
\begin{figure}[tbp]
 \includegraphics[width=0.5\textwidth]{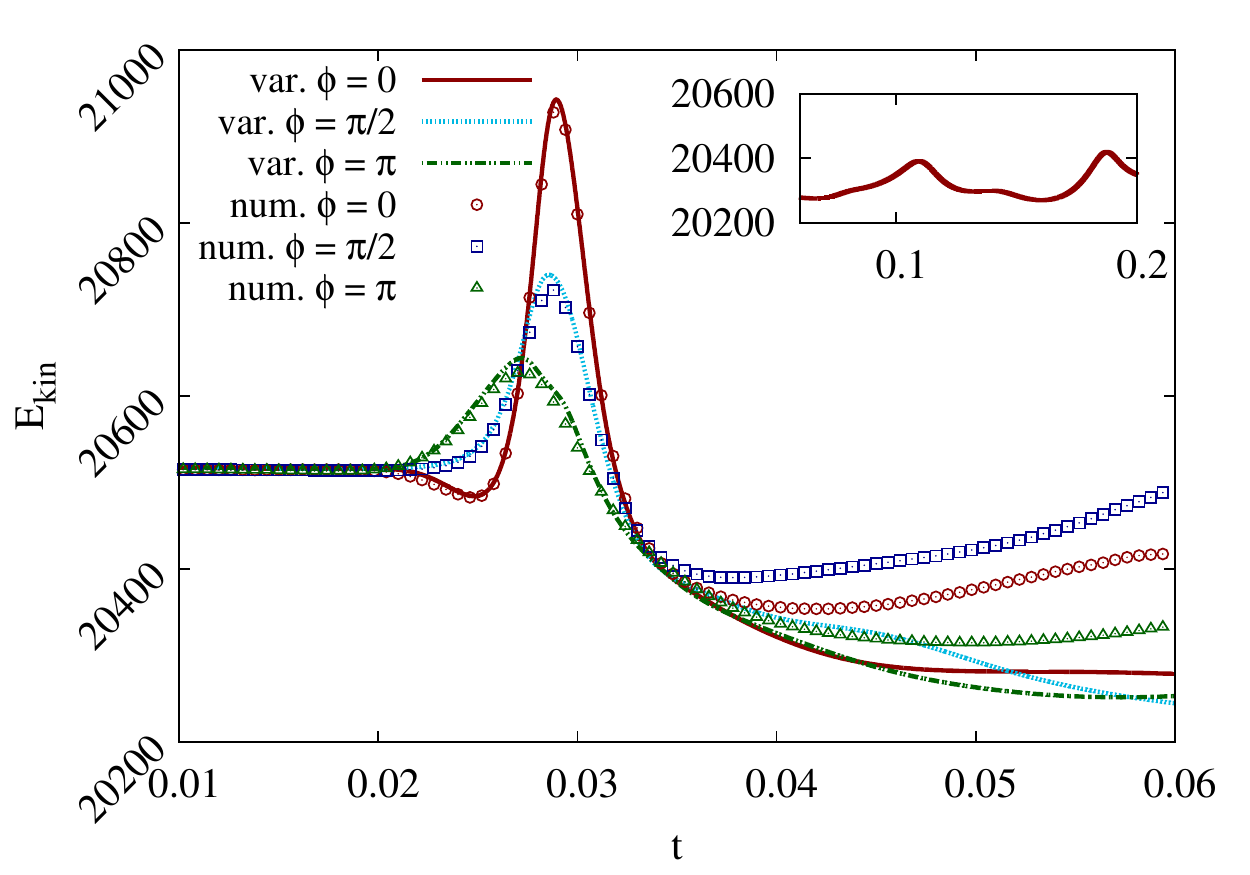}
 \caption{(Color online) Kinetic energy as a function of time for the
   collisions with angular momentum (Fig.\
   \ref{fig:collision_angMom_all}). The dots show the numerical
   results, the lines show the results obtained by the variational
   calculations. The kinetic energy increases while the condensates
   merge. After the split up, the condensates have a lower kinetic
   energy than before, indicative of a transfer from kinetic to
   internal energy, thus resulting in excitation of the condensates.
   The inset shows the kinetic energy obtained by the variational
   calculations for large timescales. The oscillatory behavior
   indicates the excitation of the solitons. Note the larger kinetic
   energy obtained by the grid calculations shortly after the split
   up. This is due to the finite grid size, which manifests itself in
   oscillations of the wave function's amplitude for large times.}
\label{fig:energy_kinetic}
\end{figure}
\begin{figure}[tbp]
  \includegraphics[width=\columnwidth]{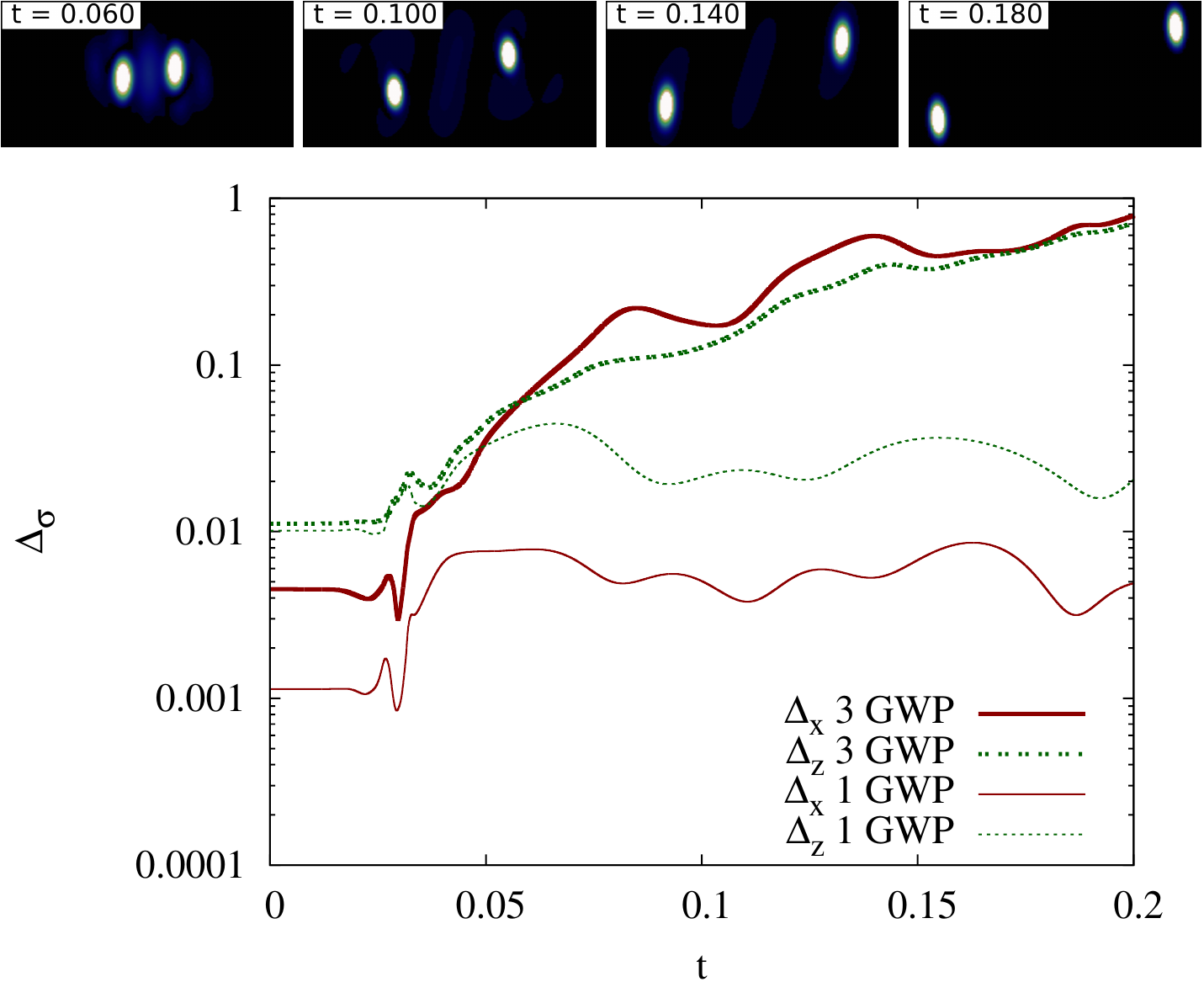}
  \caption{(Color online) Variance $\Delta_\sigma$ of a single soliton
    (right soliton in the absorption images in the upper panel) as a
    function of time. The collision occurs at $t\approx 0.03$. The
    thick solid line and the double-dashed line show the $\Delta_x$
    and $\Delta_z$ variance of three GWPs, respectively. The thin
    solid line and the normal dashed line show the $\Delta_x$ and
    $\Delta_z$ variance of the dominant GWP $g^0$.
   }
   \label{fig:variance}
 \end{figure}
The case with angular momentum is suited best to show how the transfer
of kinetic energy affects the spatial distribution of the condensate.
In Fig.~\ref{fig:energy_kinetic} we show the kinetic energy as a
function of time for the collisions with angular momentum. Comparing
the curves in Fig.~\ref{fig:energy_kinetic} with the absorption images
in Fig.~\ref{fig:collision_angMom_all}, it is obvious that a larger
transfer of kinetic energy implies a larger condensate at $t = 0.049$.
The slightly smaller transfer observed at the end of the
full-numerical calculations (this leads to a larger extension of the
solitons after the collision c.f.\ Fig.~\ref{fig:collision_angMom_all})
originates from finite grid sizes and thus has no physical meaning.
Variational calculations show an oscillation of the kinetic energy for
large timescales, which corresponds to the excitation of the solitons.

The amount of kinetic energy transferred into internal energy of the
solitons depends on the overlap of the wave functions during the
collision process. A large overlap of the solitons enhances the
nonlinear coupling in the GPE as $|\Psi(\vec{x},t)|^2$ increases and a
small one diminishes the coupling. This can be seen best in
Fig.~\ref{fig:results_10} (right column) where the destructive
interference for the calculation with phase difference $\phi=\pi$
leads to $|\Psi(\vec{0},t)|^2 = 0$. For the corresponding calculation
with nonzero angular momentum (Fig.~\ref{fig:collision_angMom_all}c)
we find $|\Psi(\vec{0},t)|^2 = 0$, too. However, the ring-like
structure increases the overlap during the collision.

In Fig.~\ref{fig:variance} the variance $\Delta_\sigma =
\left<\sigma^2 \right> - \left<\sigma \right>^2$ with $\sigma=x,z$ is
plotted as a function of time. The variance has been calculated for
the three GWPs representing the solitons on the left-hand side in the
starting configuration and for the GWP which has the largest amplitude
after the collision process. This dominant GWP $g^0$ shows oscillatory
behavior while the other GWPs with much smaller amplitudes describe
particles leaving the soliton. This effect can hardly be seen in the
absorption images in the upper panel of Fig.~\ref{fig:variance}.
However, the absorption images show that a soliton still exists,
although this would be difficult to see in an actual experiment due to
the very long time scale.

We have also performed simulations with smaller and larger momenta of
the solitons. The former case leads to one merged condensate which
does not split up again after the collision but shows oscillatory
behavior. This is very similar to the collision presented in
\cite{Pedri2005}. In the latter case the wavelength of the
interference pattern is smaller and becomes more pronounced. Note that
grid calculations with high momenta are problematic, because the
condensates quickly reach the edge of the grid. An approach with a
variational ansatz is better suited to analyze these scenarios.

\section{Conclusion}
\label{sec:conclusion}

We have studied the collisions of anisotropic two-dimensional bright
solitons in dipolar Bose-Einstein condensates both with a
fully-numerical ansatz and a time-dependent variational principle with
coupled Gaussians. The calculations presented show that the collision
process leads to an energy transfer from kinetic energy to ``inner''
energy of the solitons which leads to excited solitons with larger
extent. The absorption images show very good qualitative agreement of
the results gained by the two different methods.

The advantages of the grid calculations are the simplicity of the
numerical scheme (although the implementation for the massively
parallel computation requires some effort), the freedom in describing
all different shapes of wave functions, and the numerical stability of the
method. The advantages of the variational calculations are the much
smaller numerical effort, enabling one to run long calculations on
standard PCs, the independence of finite grid size, and the small
amount of parameters to be saved.

Both methods can be used to simulate the time-dependent GPE,
supporting each other mutually. One further application would be the
inclusion of additional external potentials such as optical lattices and
the comparison of the methods in such scenarios. Our results should
stimulate experimental efforts to study the collisions of 2D
anisotropic solitons.

\section{Acknowledgements}
We thank Boris Malomed for valuable discussions. This work was
supported by Deutsche Forschungsgemeinschaft. R.E.\ is grateful for
support from the Landes\-graduierten\-f\"orderung of the Land
Baden-W\"urttemberg.

\appendix{}

\section{Transformation to $C\! B$-variables}
\label{sec:transformation-c-b}
The direct numerical integration of Eq.~(\ref{eq:var_EOM})
leads to numerical difficulties \cite{heller:4979}. These can be dealt
with by the introduction of two auxiliary matrices $B$ and $C$. With
$A=BC^{-1}$ the equations of motion for the width matrices can be
written as
\begin{align}
  \dot A^k &= -4\ii \left( A^k \right)^2 + \ii V_2^k \,,\\
  \dot A^k &= \dot B^k
  \left(
    C^k
  \right)^{-1} - B^k
  \left(
    C^k
  \right)^{-2} \dot C^k \,,
\end{align}
where $C$ and $B$ are $3 \times 3$ complex matrices. Omitting the
index $k$ we obtain from these equations
\begin{align}
  B^{-1} \dot B C^{-1} - C^{-2} \dot C &= 
  -4 \ii B^{-1} A^2 + \ii B^{-1} V_2 \\
  \Rightarrow \quad 
  B^{-1} \dot B - C^{-2} \dot C C &=
  -4 \ii C^{-1} B + \ii B^{-1} V_2 C\,.
\end{align}
By comparison we yield the equations of motion for $C$ and $B$
\begin{align}
  \label{eq:42}
  \dot B^k &= \ii V_2^k C^k \,,\\
  \dot C^k &= 4 \ii B^k\,.
\end{align}
The reduction (\ref{eq:variational_reduction}) can be done for those
matrices, too. Note however, that the matrices $B$ and $C$ do not
preserve the same symmetry as the matrices $A$ which are complex
symmetric. Therefore, all five complex entries in $B$ and $C$ have to
be integrated.

\section{Solution of the dipolar integral}
\label{sec:vari-solut-dipol}

The calculation of the dipolar integrals needed in the TDVP
$\matrixel{\Psi}{\alpha^n\beta^m V_{\mathrm{d}}}{\Psi}$ with
$\alpha,\beta=x,y,z$ and $0 \leq n+m \leq 2$ is shown here for the
simplest case $n=m=0$. The other integrals are calculated
analogously. We start from the six-dimensional non-local integral
\begin{align}
  \label{eq:var_dipolar_six-dim}
  \matrixel{\Psi}{V_\mathrm{d}}{\Psi} =
    \sum\limits_{l,k,j,i} &\iint \dd^3r \dd^3r'\;
    {g^l}^*(\vec r)  {g^j}^*(\vec r')
    {g^i}(\vec r')  {g^k}(\vec r)\nonumber \\
    & \times \left(
      1-\frac{3(z-z')^2}{\abs{\vec r - \vec r'}^2}
    \right)
    \frac{1}{\abs{\vec r -\vec r'}^3}  \,.
\end{align}
By the use of the convolution theorem of Fourier analysis we can
evaluate one of the three-dimensional integrals directly, while the
inverse Fourier transform
\begin{align}
  \label{eq:var_dipole_int_fourier_left}
  \matrixel{\Psi}{V_\mathrm{d}}{\Psi} &=
\frac{1}{6\pi^2}\sum\limits_{l,k,j,i} I_0^{kl}I_0^{ij}\nonumber\\
&\times\int \dd^3k\;
\exp
\left\{-
  \frac{1}{4}
  \vec k^T
    \bar A^{klij}
  \vec k
  + \frac{1}{2} \ii  \left(
    \bar{\vec p}^{klij}
  \right)^T
  \vec k
\right\}\nonumber\\
&\times   \left(\frac{3k_z^2}{\vec k^2}-1 \right) \,,
\end{align}
remains to be done. Here $I_0^{kl}$ denotes the overlap integral of
the Gaussian functions $k$ and $l$, and we have used the abbreviations
\begin{align}
  \bar A^{klij} &= (A^{kl})^{-1} +(A^{ij})^{-1}   \label{eq:Abar}\,, \\
  \bar{\vec p}^{klij} &= (A^{ij})^{-1} \vec p^{ij} - (A^{kl})^{-1}
  \vec p^{kl}\,, \label{eq:pbar}
\end{align}
with $A^{kl}=A^k+{A^{l}}^*$ and ${\vec p}^{kl} = {\vec p}^k + {\vec
  p^l}^*$ and analogously for $i$ and $j$. The integral
(\ref{eq:var_dipole_int_fourier_left}) can be split in two parts, one
leading to a shift in the scattering length (this is the short-range
part of the DDI) and a second part
$\matrixel{\Psi}{V_\mathrm{d,eff}}{\Psi} =\sum_{l,k,j,i}
I_0^{kl}I_0^{ij} J_2^{klij} $. After a principal component analysis of
the exponential in Eq.~(\ref{eq:var_dipole_int_fourier_left}) the
analytical integration in $k_y$-direction is possible when we make use
of Eq.~(\ref{eq:variational_reduction}). The remaining result reads
\begin{align}
  \label{eq:var_J2result}
J_2^{klij} &=  
\frac{1}{4\pi} \int\limits_0^{\infty}\dd \rho\; w
\left( \ii
\sqrt{\bar A_y^{klij}} \frac{\rho}{2}
\right)\rho^2 \ee{-\frac{1}{8} \left(
    \bar A_x^{klij} + \bar A_z^{klij}
  \right)    \rho^2}
\nonumber\\
&\times
\sum\limits_{\substack{\pm x_\mathrm{c} \\ \pm x_\mathrm{s}}}\;\int\limits_{-1}^{1} \!
\dd x\;\frac{(\pm c_1 x_\mathrm{c} \pm c_0 x_\mathrm{s})^2}{\sqrt{1-x^2}} \nonumber \\
&\times\ee{-\frac{1}{8}
  \left(
    \bar A_x^{klij} - \bar A_z^{klij}
  \right) \rho^2 x + \frac{\ii}{2}
  \left(
    \pm\bar p_x^{klij} \rho x_\mathrm{c} \pm \bar p_z^{klij} \rho x_\mathrm{s}
  \right)}\,,
\end{align}
with $x_c = \sqrt{(1+x)/2} ,\; x_s =\sqrt{(1-x)/2}$, the coefficients
$c_0,\,c_1$ of the rotation matrix from the principal component
analysis and the Faddeeva function $w(z)=\ee{-z^2} \erfc (-\ii z)$.
The numerical evaluation of this integral can efficiently be performed by
a Taylor expansion of the Faddeeva function for which the single terms can
be obtained by a recursion formula and using a Chebyshev quadrature
for the $x$-integration. To improve the result we apply a
Pad\'e-approximation to the Taylor series.

The numerical integration of the dipolar integrals is the crucial part
in this method. Dependent on the number of Gaussian functions $N$
there is a total number of $C_\mathrm{num} = (N^4+N^2-2N)/4$
integrals to be calculated numerically and $C_\mathrm{elliptic} =
N(N+1)/2$ 
which can be expressed in terms of elliptic integrals.


\begin{thebibliography}{24}%
\makeatletter
\providecommand \@ifxundefined [1]{%
 \@ifx{#1\undefined}
}%
\providecommand \@ifnum [1]{%
 \ifnum #1\expandafter \@firstoftwo
 \else \expandafter \@secondoftwo
 \fi
}%
\providecommand \@ifx [1]{%
 \ifx #1\expandafter \@firstoftwo
 \else \expandafter \@secondoftwo
 \fi
}%
\providecommand \natexlab [1]{#1}%
\providecommand \enquote  [1]{``#1''}%
\providecommand \bibnamefont  [1]{#1}%
\providecommand \bibfnamefont [1]{#1}%
\providecommand \citenamefont [1]{#1}%
\providecommand \href@noop [0]{\@secondoftwo}%
\providecommand \href [0]{\begingroup \@sanitize@url \@href}%
\providecommand \@href[1]{\@@startlink{#1}\@@href}%
\providecommand \@@href[1]{\endgroup#1\@@endlink}%
\providecommand \@sanitize@url [0]{\catcode `\\12\catcode `\$12\catcode
  `\&12\catcode `\#12\catcode `\^12\catcode `\_12\catcode `\%12\relax}%
\providecommand \@@startlink[1]{}%
\providecommand \@@endlink[0]{}%
\providecommand \url  [0]{\begingroup\@sanitize@url \@url }%
\providecommand \@url [1]{\endgroup\@href {#1}{\urlprefix }}%
\providecommand \urlprefix  [0]{URL }%
\providecommand \Eprint [0]{\href }%
\providecommand \doibase [0]{http://dx.doi.org/}%
\providecommand \selectlanguage [0]{\@gobble}%
\providecommand \bibinfo  [0]{\@secondoftwo}%
\providecommand \bibfield  [0]{\@secondoftwo}%
\providecommand \translation [1]{[#1]}%
\providecommand \BibitemOpen [0]{}%
\providecommand \bibitemStop [0]{}%
\providecommand \bibitemNoStop [0]{.\EOS\space}%
\providecommand \EOS [0]{\spacefactor3000\relax}%
\providecommand \BibitemShut  [1]{\csname bibitem#1\endcsname}%
\let\auto@bib@innerbib\@empty
\bibitem [{\citenamefont {Griesmaier}\ \emph {et~al.}(2005)\citenamefont
  {Griesmaier}, \citenamefont {Werner}, \citenamefont {Hensler}, \citenamefont
  {Stuhler},\ and\ \citenamefont {Pfau}}]{Griesmaier2005}%
  \BibitemOpen
  \bibfield  {author} {\bibinfo {author} {\bibfnamefont {A.}~\bibnamefont
  {Griesmaier}}, \bibinfo {author} {\bibfnamefont {J.}~\bibnamefont {Werner}},
  \bibinfo {author} {\bibfnamefont {S.}~\bibnamefont {Hensler}}, \bibinfo
  {author} {\bibfnamefont {J.}~\bibnamefont {Stuhler}}, \ and\ \bibinfo
  {author} {\bibfnamefont {T.}~\bibnamefont {Pfau}},\ }\href@noop {} {\bibfield
   {journal} {\bibinfo  {journal} {{Phys. Rev. Lett.}}\ }\textbf {\bibinfo
  {volume} {94}},\ \bibinfo {pages} {160401} (\bibinfo {year}
  {2005})}\BibitemShut {NoStop}%
\bibitem [{\citenamefont {Beaufils}\ \emph {et~al.}(2008)\citenamefont
  {Beaufils}, \citenamefont {Chicireanu}, \citenamefont {Zanon}, \citenamefont
  {Laburthe-Tolra}, \citenamefont {Mar{\'e}chal}, \citenamefont {Vernac},
  \citenamefont {Keller},\ and\ \citenamefont {Gorceix}}]{Beaufils2008}%
  \BibitemOpen
  \bibfield  {author} {\bibinfo {author} {\bibfnamefont {Q.}~\bibnamefont
  {Beaufils}}, \bibinfo {author} {\bibfnamefont {R.}~\bibnamefont
  {Chicireanu}}, \bibinfo {author} {\bibfnamefont {T.}~\bibnamefont {Zanon}},
  \bibinfo {author} {\bibfnamefont {B.}~\bibnamefont {Laburthe-Tolra}},
  \bibinfo {author} {\bibfnamefont {E.}~\bibnamefont {Mar{\'e}chal}}, \bibinfo
  {author} {\bibfnamefont {L.}~\bibnamefont {Vernac}}, \bibinfo {author}
  {\bibfnamefont {J.-C.}\ \bibnamefont {Keller}}, \ and\ \bibinfo {author}
  {\bibfnamefont {O.}~\bibnamefont {Gorceix}},\ }\href {\doibase
  10.1103/PhysRevA.77.061601} {\bibfield  {journal} {\bibinfo  {journal} {Phys.
  Rev. A}\ }\textbf {\bibinfo {volume} {77}},\ \bibinfo {pages} {061601(R)}
  (\bibinfo {year} {2008})}\BibitemShut {NoStop}%
\bibitem [{\citenamefont {Lu}\ \emph {et~al.}(2010)\citenamefont {Lu},
  \citenamefont {Youn},\ and\ \citenamefont {Lev}}]{Lu2010}%
  \BibitemOpen
  \bibfield  {author} {\bibinfo {author} {\bibfnamefont {M.}~\bibnamefont
  {Lu}}, \bibinfo {author} {\bibfnamefont {S.~H.}\ \bibnamefont {Youn}}, \ and\
  \bibinfo {author} {\bibfnamefont {B.~L.}\ \bibnamefont {Lev}},\ }\href@noop
  {} {\bibfield  {journal} {\bibinfo  {journal} {{Phys. Rev. Lett.}}\ }\textbf
  {\bibinfo {volume} {104}},\ \bibinfo {pages} {063001} (\bibinfo {year}
  {2010})}\BibitemShut {NoStop}%
\bibitem [{\citenamefont {Lu}\ \emph {et~al.}(2011)\citenamefont {Lu},
  \citenamefont {Burdick}, \citenamefont {Youn},\ and\ \citenamefont
  {Lev}}]{Lu2011_2}%
  \BibitemOpen
  \bibfield  {author} {\bibinfo {author} {\bibfnamefont {M.}~\bibnamefont
  {Lu}}, \bibinfo {author} {\bibfnamefont {N.~Q.}\ \bibnamefont {Burdick}},
  \bibinfo {author} {\bibfnamefont {S.~H.}\ \bibnamefont {Youn}}, \ and\
  \bibinfo {author} {\bibfnamefont {B.~L.}\ \bibnamefont {Lev}},\ }\href@noop
  {} {\bibfield  {journal} {\bibinfo  {journal} {Phys. Rev. Lett.}\ }\textbf
  {\bibinfo {volume} {107}},\ \bibinfo {pages} {190401} (\bibinfo {year}
  {2011})}\BibitemShut {NoStop}%
\bibitem [{\citenamefont {Aikawa}\ \emph {et~al.}(2012)\citenamefont {Aikawa},
  \citenamefont {Frisch}, \citenamefont {Mark}, \citenamefont {Baier},
  \citenamefont {Rietzler}, \citenamefont {Grimm},\ and\ \citenamefont
  {Ferlaino}}]{Aikawa_2012}%
  \BibitemOpen
  \bibfield  {author} {\bibinfo {author} {\bibfnamefont {K.}~\bibnamefont
  {Aikawa}}, \bibinfo {author} {\bibfnamefont {A.}~\bibnamefont {Frisch}},
  \bibinfo {author} {\bibfnamefont {M.}~\bibnamefont {Mark}}, \bibinfo {author}
  {\bibfnamefont {S.}~\bibnamefont {Baier}}, \bibinfo {author} {\bibfnamefont
  {A.}~\bibnamefont {Rietzler}}, \bibinfo {author} {\bibfnamefont
  {R.}~\bibnamefont {Grimm}}, \ and\ \bibinfo {author} {\bibfnamefont
  {F.}~\bibnamefont {Ferlaino}},\ }\href {\doibase
  10.1103/PhysRevLett.108.210401} {\bibfield  {journal} {\bibinfo  {journal}
  {Phys. Rev. Lett.}\ }\textbf {\bibinfo {volume} {108}},\ \bibinfo {pages}
  {210401} (\bibinfo {year} {2012})}\BibitemShut {NoStop}%
\bibitem [{\citenamefont {Ni}\ \emph {et~al.}(2008)\citenamefont {Ni},
  \citenamefont {Ospelkaus}, \citenamefont {de~Miranda}, \citenamefont {Pe'er},
  \citenamefont {Neyenhuis}, \citenamefont {Zirbel}, \citenamefont
  {Kotochigova}, \citenamefont {Julienne}, \citenamefont {Jin},\ and\
  \citenamefont {Ye}}]{Ni_2008}%
  \BibitemOpen
  \bibfield  {author} {\bibinfo {author} {\bibfnamefont {K.-K.}\ \bibnamefont
  {Ni}}, \bibinfo {author} {\bibfnamefont {S.}~\bibnamefont {Ospelkaus}},
  \bibinfo {author} {\bibfnamefont {M.~H.~G.}\ \bibnamefont {de~Miranda}},
  \bibinfo {author} {\bibfnamefont {A.}~\bibnamefont {Pe'er}}, \bibinfo
  {author} {\bibfnamefont {B.}~\bibnamefont {Neyenhuis}}, \bibinfo {author}
  {\bibfnamefont {J.~J.}\ \bibnamefont {Zirbel}}, \bibinfo {author}
  {\bibfnamefont {S.}~\bibnamefont {Kotochigova}}, \bibinfo {author}
  {\bibfnamefont {P.~S.}\ \bibnamefont {Julienne}}, \bibinfo {author}
  {\bibfnamefont {D.~S.}\ \bibnamefont {Jin}}, \ and\ \bibinfo {author}
  {\bibfnamefont {J.}~\bibnamefont {Ye}},\ }\href@noop {} {\bibfield  {journal}
  {\bibinfo  {journal} {Science}\ }\textbf {\bibinfo {volume} {332}},\ \bibinfo
  {pages} {231} (\bibinfo {year} {2008})}\BibitemShut {NoStop}%
\bibitem [{\citenamefont {Lahaye}\ \emph {et~al.}(2009)\citenamefont {Lahaye},
  \citenamefont {Menotti}, \citenamefont {Santos}, \citenamefont {Lewenstein},\
  and\ \citenamefont {Pfau}}]{Lahaye2009}%
  \BibitemOpen
  \bibfield  {author} {\bibinfo {author} {\bibfnamefont {T.}~\bibnamefont
  {Lahaye}}, \bibinfo {author} {\bibfnamefont {C.}~\bibnamefont {Menotti}},
  \bibinfo {author} {\bibfnamefont {L.}~\bibnamefont {Santos}}, \bibinfo
  {author} {\bibfnamefont {M.}~\bibnamefont {Lewenstein}}, \ and\ \bibinfo
  {author} {\bibfnamefont {T.}~\bibnamefont {Pfau}},\ }\href@noop {} {\bibfield
   {journal} {\bibinfo  {journal} {Rep. Progr. Phys.}\ }\textbf {\bibinfo
  {volume} {72}},\ \bibinfo {pages} {126401} (\bibinfo {year}
  {2009})}\BibitemShut {NoStop}%
\bibitem [{\citenamefont {Khaykovich}\ \emph {et~al.}(2002)\citenamefont
  {Khaykovich}, \citenamefont {Schreck}, \citenamefont {Ferrari}, \citenamefont
  {Bourdel}, \citenamefont {Cubizolles}, \citenamefont {Carr},\ and\
  \citenamefont {Castin}}]{Khaykovich2002}%
  \BibitemOpen
  \bibfield  {author} {\bibinfo {author} {\bibfnamefont {L.}~\bibnamefont
  {Khaykovich}}, \bibinfo {author} {\bibfnamefont {F.}~\bibnamefont {Schreck}},
  \bibinfo {author} {\bibfnamefont {G.}~\bibnamefont {Ferrari}}, \bibinfo
  {author} {\bibfnamefont {T.}~\bibnamefont {Bourdel}}, \bibinfo {author}
  {\bibfnamefont {J.}~\bibnamefont {Cubizolles}}, \bibinfo {author}
  {\bibfnamefont {L.~D.}\ \bibnamefont {Carr}}, \ and\ \bibinfo {author}
  {\bibfnamefont {Y.}~\bibnamefont {Castin}},\ }\href@noop {} {\bibfield
  {journal} {\bibinfo  {journal} {Science}\ }\textbf {\bibinfo {volume}
  {296}},\ \bibinfo {pages} {1290} (\bibinfo {year} {2002})}\BibitemShut
  {NoStop}%
\bibitem [{\citenamefont {Strecker}\ \emph {et~al.}(2002)\citenamefont
  {Strecker}, \citenamefont {Partridge}, \citenamefont {Truscott},\ and\
  \citenamefont {Hulet}}]{Strecker2002}%
  \BibitemOpen
  \bibfield  {author} {\bibinfo {author} {\bibfnamefont {K.~E.}\ \bibnamefont
  {Strecker}}, \bibinfo {author} {\bibfnamefont {G.~B.}\ \bibnamefont
  {Partridge}}, \bibinfo {author} {\bibfnamefont {A.~G.}\ \bibnamefont
  {Truscott}}, \ and\ \bibinfo {author} {\bibfnamefont {R.~G.}\ \bibnamefont
  {Hulet}},\ }\href@noop {} {\bibfield  {journal} {\bibinfo  {journal}
  {Nature}\ }\textbf {\bibinfo {volume} {417}},\ \bibinfo {pages} {150}
  (\bibinfo {year} {2002})}\BibitemShut {NoStop}%
\bibitem [{\citenamefont {Tikhonenkov}\ \emph {et~al.}(2008)\citenamefont
  {Tikhonenkov}, \citenamefont {Malomed},\ and\ \citenamefont
  {Vardi}}]{Tikhonenkov2008_2}%
  \BibitemOpen
  \bibfield  {author} {\bibinfo {author} {\bibfnamefont {I.}~\bibnamefont
  {Tikhonenkov}}, \bibinfo {author} {\bibfnamefont {B.~A.}\ \bibnamefont
  {Malomed}}, \ and\ \bibinfo {author} {\bibfnamefont {A.}~\bibnamefont
  {Vardi}},\ }\href@noop {} {\bibfield  {journal} {\bibinfo  {journal} {{Phys.
  Rev. Lett.}}\ }\textbf {\bibinfo {volume} {100}},\ \bibinfo {pages} {090406}
  (\bibinfo {year} {2008})}\BibitemShut {NoStop}%
\bibitem [{\citenamefont {K{\"o}berle}\ \emph {et~al.}(2012)\citenamefont
  {K{\"o}berle}, \citenamefont {Zajec}, \citenamefont {Wunner},\ and\
  \citenamefont {Malomed}}]{Koeberle2012}%
  \BibitemOpen
  \bibfield  {author} {\bibinfo {author} {\bibfnamefont {P.}~\bibnamefont
  {K{\"o}berle}}, \bibinfo {author} {\bibfnamefont {D.}~\bibnamefont {Zajec}},
  \bibinfo {author} {\bibfnamefont {G.}~\bibnamefont {Wunner}}, \ and\ \bibinfo
  {author} {\bibfnamefont {B.~A.}\ \bibnamefont {Malomed}},\ }\href@noop {}
  {\bibfield  {journal} {\bibinfo  {journal} {Physical Review A}\ }\textbf
  {\bibinfo {volume} {85}},\ \bibinfo {pages} {023630} (\bibinfo {year}
  {2012})}\BibitemShut {NoStop}%
\bibitem [{\citenamefont {Eichler}\ \emph {et~al.}(2011)\citenamefont
  {Eichler}, \citenamefont {Main},\ and\ \citenamefont {Wunner}}]{Eichler2011}%
  \BibitemOpen
  \bibfield  {author} {\bibinfo {author} {\bibfnamefont {R.}~\bibnamefont
  {Eichler}}, \bibinfo {author} {\bibfnamefont {J.}~\bibnamefont {Main}}, \
  and\ \bibinfo {author} {\bibfnamefont {G.}~\bibnamefont {Wunner}},\
  }\href@noop {} {\bibfield  {journal} {\bibinfo  {journal} {{Phys. Rev. A}}\
  }\textbf {\bibinfo {volume} {83}},\ \bibinfo {pages} {053604} (\bibinfo
  {year} {2011})}\BibitemShut {NoStop}%
\bibitem [{\citenamefont {Adhikari}\ and\ \citenamefont
  {Muruganandam}(2012)}]{Adhikari_2012}%
  \BibitemOpen
  \bibfield  {author} {\bibinfo {author} {\bibfnamefont {S.~K.}\ \bibnamefont
  {Adhikari}}\ and\ \bibinfo {author} {\bibfnamefont {P.}~\bibnamefont
  {Muruganandam}},\ }\href@noop {} {\bibfield  {journal} {\bibinfo  {journal}
  {J. Phys. B: At. Mol. Opt. Phys.}\ }\textbf {\bibinfo {volume} {45}},\
  \bibinfo {pages} {045301} (\bibinfo {year} {2012})}\BibitemShut {NoStop}%
\bibitem [{\citenamefont {Pedri}\ and\ \citenamefont
  {Santos}(2005)}]{Pedri2005}%
  \BibitemOpen
  \bibfield  {author} {\bibinfo {author} {\bibfnamefont {P.}~\bibnamefont
  {Pedri}}\ and\ \bibinfo {author} {\bibfnamefont {L.}~\bibnamefont {Santos}},\
  }\href@noop {} {\bibfield  {journal} {\bibinfo  {journal} {{Phys. Rev.
  Lett.}}\ }\textbf {\bibinfo {volume} {95}},\ \bibinfo {pages} {200404}
  (\bibinfo {year} {2005})}\BibitemShut {NoStop}%
\bibitem [{\citenamefont {Giovanazzi}\ \emph {et~al.}(2002)\citenamefont
  {Giovanazzi}, \citenamefont {G{\"o}rlitz},\ and\ \citenamefont
  {Pfau}}]{Giovanazzi2002}%
  \BibitemOpen
  \bibfield  {author} {\bibinfo {author} {\bibfnamefont {S.}~\bibnamefont
  {Giovanazzi}}, \bibinfo {author} {\bibfnamefont {A.}~\bibnamefont
  {G{\"o}rlitz}}, \ and\ \bibinfo {author} {\bibfnamefont {T.}~\bibnamefont
  {Pfau}},\ }\href@noop {} {\bibfield  {journal} {\bibinfo  {journal} {Phys.
  Rev. Lett.}\ }\textbf {\bibinfo {volume} {89}},\ \bibinfo {pages} {130401}
  (\bibinfo {year} {2002})}\BibitemShut {NoStop}%
\bibitem [{\citenamefont {Young}\ \emph {et~al.}(2011)\citenamefont {Young},
  \citenamefont {Muruganandam},\ and\ \citenamefont {Adhikari}}]{Young2011}%
  \BibitemOpen
  \bibfield  {author} {\bibinfo {author} {\bibfnamefont {L.~E.}\ \bibnamefont
  {Young}}, \bibinfo {author} {\bibfnamefont {P.}~\bibnamefont {Muruganandam}},
  \ and\ \bibinfo {author} {\bibfnamefont {S.~K.}\ \bibnamefont {Adhikari}},\
  }\href@noop {} {\bibfield  {journal} {\bibinfo  {journal} {J. Phys. B: At.
  Mol. Opt. Phys.}\ }\textbf {\bibinfo {volume} {44}},\ \bibinfo {pages}
  {101001} (\bibinfo {year} {2011})}\BibitemShut {NoStop}%
\bibitem [{\citenamefont {K{\"o}berle}\ \emph {et~al.}(2009)\citenamefont
  {K{\"o}berle}, \citenamefont {Cartarius}, \citenamefont {Fab\v{c}i\v{c}},
  \citenamefont {Main},\ and\ \citenamefont {Wunner}}]{Koeberle09a}%
  \BibitemOpen
  \bibfield  {author} {\bibinfo {author} {\bibfnamefont {P.}~\bibnamefont
  {K{\"o}berle}}, \bibinfo {author} {\bibfnamefont {H.}~\bibnamefont
  {Cartarius}}, \bibinfo {author} {\bibfnamefont {T.}~\bibnamefont
  {Fab\v{c}i\v{c}}}, \bibinfo {author} {\bibfnamefont {J.}~\bibnamefont
  {Main}}, \ and\ \bibinfo {author} {\bibfnamefont {G.}~\bibnamefont
  {Wunner}},\ }\href@noop {} {\bibfield  {journal} {\bibinfo  {journal} {New
  Journal of Physics}\ }\textbf {\bibinfo {volume} {11}},\ \bibinfo {pages}
  {023017} (\bibinfo {year} {2009})}\BibitemShut {NoStop}%
\bibitem [{\citenamefont {Chin}\ \emph {et~al.}(2010)\citenamefont {Chin},
  \citenamefont {Grimm}, \citenamefont {Julienne},\ and\ \citenamefont
  {Tiesinga}}]{Chin2010}%
  \BibitemOpen
  \bibfield  {author} {\bibinfo {author} {\bibfnamefont {C.}~\bibnamefont
  {Chin}}, \bibinfo {author} {\bibfnamefont {R.}~\bibnamefont {Grimm}},
  \bibinfo {author} {\bibfnamefont {P.}~\bibnamefont {Julienne}}, \ and\
  \bibinfo {author} {\bibfnamefont {E.}~\bibnamefont {Tiesinga}},\ }\href@noop
  {} {\bibfield  {journal} {\bibinfo  {journal} {Rev. Mod. Phys.}\ }\textbf
  {\bibinfo {volume} {82}},\ \bibinfo {pages} {1225} (\bibinfo {year}
  {2010})}\BibitemShut {NoStop}%
\bibitem [{\citenamefont {Feit}\ \emph {et~al.}(1982)\citenamefont {Feit},
  \citenamefont {Fleck}, \citenamefont {Jr.},\ and\ \citenamefont
  {Steiger}}]{Feit_1982}%
  \BibitemOpen
  \bibfield  {author} {\bibinfo {author} {\bibfnamefont {M.~D.}\ \bibnamefont
  {Feit}}, \bibinfo {author} {\bibfnamefont {J.~A.}\ \bibnamefont {Fleck}},
  \bibinfo {author} {\bibnamefont {Jr.}}, \ and\ \bibinfo {author}
  {\bibfnamefont {A.}~\bibnamefont {Steiger}},\ }\href@noop {} {\bibfield
  {journal} {\bibinfo  {journal} {J. Comp. Phys.}\ }\textbf {\bibinfo {volume}
  {47}},\ \bibinfo {pages} {412} (\bibinfo {year} {1982})}\BibitemShut
  {NoStop}%
\bibitem [{\citenamefont {Rau}\ \emph {et~al.}(2010{\natexlab{a}})\citenamefont
  {Rau}, \citenamefont {Main}, \citenamefont {K{\"o}berle},\ and\ \citenamefont
  {Wunner}}]{Rau10}%
  \BibitemOpen
  \bibfield  {author} {\bibinfo {author} {\bibfnamefont {S.}~\bibnamefont
  {Rau}}, \bibinfo {author} {\bibfnamefont {J.}~\bibnamefont {Main}}, \bibinfo
  {author} {\bibfnamefont {P.}~\bibnamefont {K{\"o}berle}}, \ and\ \bibinfo
  {author} {\bibfnamefont {G.}~\bibnamefont {Wunner}},\ }\href {\doibase
  10.1103/PhysRevA.81.031605} {\bibfield  {journal} {\bibinfo  {journal} {Phys.
  Rev. A}\ }\textbf {\bibinfo {volume} {81}},\ \bibinfo {pages} {031605(R)}
  (\bibinfo {year} {2010}{\natexlab{a}})}\BibitemShut {NoStop}%
\bibitem [{\citenamefont {McLachlan}(1964)}]{McLachlan1964a}%
  \BibitemOpen
  \bibfield  {author} {\bibinfo {author} {\bibfnamefont {A.~D.}\ \bibnamefont
  {McLachlan}},\ }\href {\doibase 10.1080/00268976400100041} {\bibfield
  {journal} {\bibinfo  {journal} {Mol. Phys.}\ }\textbf {\bibinfo {volume}
  {8}},\ \bibinfo {pages} {39} (\bibinfo {year} {1964})}\BibitemShut {NoStop}%
\bibitem [{\citenamefont {Rau}\ \emph {et~al.}(2010{\natexlab{b}})\citenamefont
  {Rau}, \citenamefont {Main},\ and\ \citenamefont {Wunner}}]{Rau10a}%
  \BibitemOpen
  \bibfield  {author} {\bibinfo {author} {\bibfnamefont {S.}~\bibnamefont
  {Rau}}, \bibinfo {author} {\bibfnamefont {J.}~\bibnamefont {Main}}, \ and\
  \bibinfo {author} {\bibfnamefont {G.}~\bibnamefont {Wunner}},\ }\href
  {\doibase 10.1103/PhysRevA.82.023610} {\bibfield  {journal} {\bibinfo
  {journal} {Phys. Rev. A}\ }\textbf {\bibinfo {volume} {82}},\ \bibinfo
  {pages} {023610} (\bibinfo {year} {2010}{\natexlab{b}})}\BibitemShut
  {NoStop}%
\bibitem [{\citenamefont {Rau}\ \emph {et~al.}(2010{\natexlab{c}})\citenamefont
  {Rau}, \citenamefont {Main}, \citenamefont {Cartarius}, \citenamefont
  {K{\"o}berle},\ and\ \citenamefont {Wunner}}]{Rau10b}%
  \BibitemOpen
  \bibfield  {author} {\bibinfo {author} {\bibfnamefont {S.}~\bibnamefont
  {Rau}}, \bibinfo {author} {\bibfnamefont {J.}~\bibnamefont {Main}}, \bibinfo
  {author} {\bibfnamefont {H.}~\bibnamefont {Cartarius}}, \bibinfo {author}
  {\bibfnamefont {P.}~\bibnamefont {K{\"o}berle}}, \ and\ \bibinfo {author}
  {\bibfnamefont {G.}~\bibnamefont {Wunner}},\ }\href {\doibase
  10.1103/PhysRevA.82.023611} {\bibfield  {journal} {\bibinfo  {journal} {Phys.
  Rev. A}\ }\textbf {\bibinfo {volume} {82}},\ \bibinfo {pages} {023611}
  (\bibinfo {year} {2010}{\natexlab{c}})}\BibitemShut {NoStop}%
\bibitem [{\citenamefont {Heller}(1976)}]{heller:4979}%
  \BibitemOpen
  \bibfield  {author} {\bibinfo {author} {\bibfnamefont {E.~J.}\ \bibnamefont
  {Heller}},\ }\href {\doibase 10.1063/1.432974} {\bibfield  {journal}
  {\bibinfo  {journal} {J. Chem. Phys.}\ }\textbf {\bibinfo {volume} {65}},\
  \bibinfo {pages} {4979} (\bibinfo {year} {1976})}\BibitemShut {NoStop}%
\end{thebibliography}

\end{document}